\documentclass[9pt,twocolumn,twoside]{osajnl}
\usepackage{amsmath}
\usepackage{mathrsfs}
\usepackage{amssymb}
\usepackage{amsfonts}
\usepackage{graphicx}
\usepackage{float}
\usepackage{graphicx}
\usepackage{subfigure}
\usepackage{epstopdf}
\journal{ol} 

\setboolean{shortarticle}{true}

\title{Iterative fluctuation ghost imaging}

\author[1]{Huan Zhao}
\author[1,**]{Xiao-Qian Wang}
\author[1]{Chao Gao}
\author[1,2]{Zhuo Yu}
\author[1]{Hong Wang}
\author[1]{Yu Wang}
\author[1]{Li-Dan Gou}
\author[1,*]{Zhi-Hai Yao}

\affil[1]{Department of Physics, Changchun University of Science and Technology, Changchun 130022, China}
\affil[2]{School of Physics and Electronics, Baicheng Normal University, Baicheng 137000, Chnia}

\affil[*]{Corresponding author: yaozhi@cust.edu.com}
\affil[**]{Corresponding author: xqwang21@163.com}

\begin{abstract}
We present a new technique, iterative fluctuation ghost imaging (IFGI) which dramatically enhances the resolution of ghost imaging (GI). It is shown that, by the fluctuation characteristics of the second-order correlation function, the imaging information with the narrower point spread function (PSF) than the original information can be got. The effects arising from the PSF and the iteration times also be discussed.

\end{abstract}

\setboolean{displaycopyright}{true}

\begin{document}

\maketitle
Ghost imaging (GI) is an imaging technique based on second-order intensity correlation\cite{pittman1995optical,bennink2002two,cheng2004incoherent,gatti2004ghost,ferri2005high,cai2005ghost,gao2019ghost,gao2017optimization,cao2005geometrical,valencia2005two,zhang2005correlated}. It has various novel advantages for practical applications compared with traditional imaging, such as resistant of atmosphere turbulence\cite{dixon2011quantum,meyers2011turbulence} and lensless iamging \cite{cheng2004incoherent,gatti2004ghost}. GI has been applied in various fields including optical lithography\cite{bentley2004nonlinear}, remote imaging\cite{meyers2008ghost},  microscopy imaging\cite{radwell2014single},  x-ray imaging\cite{pelliccia2016experimental} and imaging for an occluded object\cite{gao2019ghost}. In these applications, the spatial resolution of image is an important factor\cite{chen2017sub}. How to improve the resolution of GI is a key factor in the development of GI\cite{chen2017sub,moreau2018resolution,meng2018super,gong2012experimental}.\par
 In traditional imaging systems, the resolution is given by Rayleigh criterion\cite{ferri2005high}.It comes from the point spread function (PSF) of the imaging system. Many schemes have been proposed to enhance resolution via reducing the impact of PSF\cite{hell1994breaking,tsang2019resurgence}. In GI systems, the spatial resolution is also limited by the Rayleigh diffraction bound just as in traditional imaging, when the pixel size of the detector is much smaller than the average size of the speckles\cite{chen2017sub,moreau2018resolution}. In general, it is taken to be the full-width at half-maximum (FWHM) of the PSF, and is approximately equal to the average size of the speckles\cite{ferri2005high,chen2017sub,ferri2008longitudinal}. Some schemes to improve the spatial resolution of GI have also been proposed. Han’s group reported a two-arm microscope scheme by employing second-order intensity correlation imaging to narrow PSF\cite{zhang2009improving}. Compressed sensing GI (CSGI) reduces the effect of PSF on the imaging quality using sparsity constraints to improve the spatial resolution of GI\cite{gong2012experimental,du2012influence,shechtman2010super,chen2013sub}. Shih’s group report a super-resolution method which using the spatial-frequency filtered intensity fluctuation correlation to reduce the FWHM of PSF\cite{sprigg2016super}. The narrowing of PSF by higher-order correlation of non-Rayleigh speckle fields has been reported\cite{kuplicki2016high}. A sub-Rayleigh resolution ghost imaging experiment is performed by spatial low-pass filtering of the instantaneous intensity to narrow PSF\cite{chen2017sub}.  Li’s group used localizing and thresholding to reduces the effect of PSF in the GI system\cite{wang2019enhancement}. Other schemes to enhance the resolution of GI by reducing the effect of the PSF have also been suggested, such as preconditioned deconvolution methods\cite{tong2021preconditioned}, deep neural network constraints\cite{wang2022far}, speeded up robust features new sum of modified Laplacian (SURF-NSML)\cite{ye2022high}, high-resolution ghost imaging through complex scattering media via a temporal correction\cite{xiao2022high} and second-order cumulants ghost imaging (SCGI)\cite{zhao2022second}.\par
In this work, we propose a scheme based on traditional thermal GI. It can narrow the PSF of the GI via the fluctuation characteristics of the second-order correlation function. And further extraction of deep-seated fluctuation information via iterations, the PSF can further become narrow. Based on this effect, the scheme can significantly enhance the resolution of GI system until the resolution approaches the pixel size of the detector. We called the scheme iterative fluctuation ghost imaging (IFGI). Compare to SCGI, there is no cross-information in IFGI. Thus, IFGI has higher resolution limits than SCGI.\par
\begin{figure*}
  \centering
  \subfigure[]{\includegraphics[width=8cm]{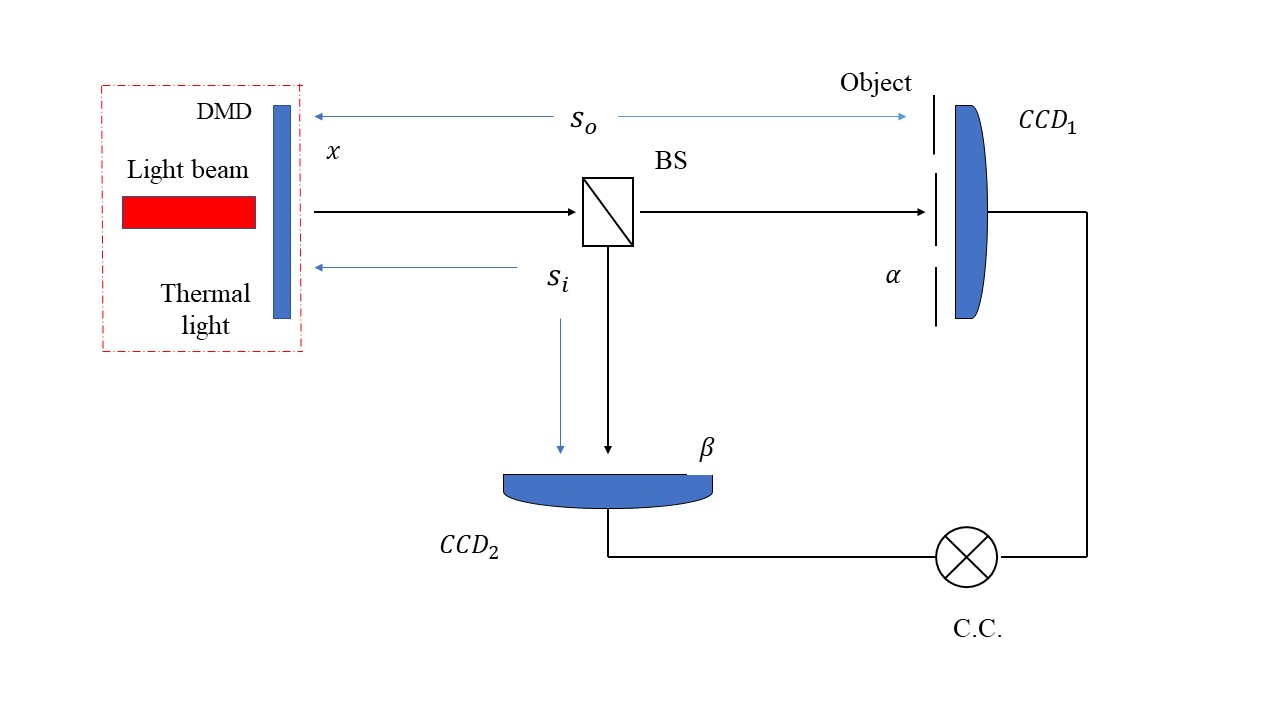}}
  \hspace{0in}
  \subfigure[]{\includegraphics[width=8cm]{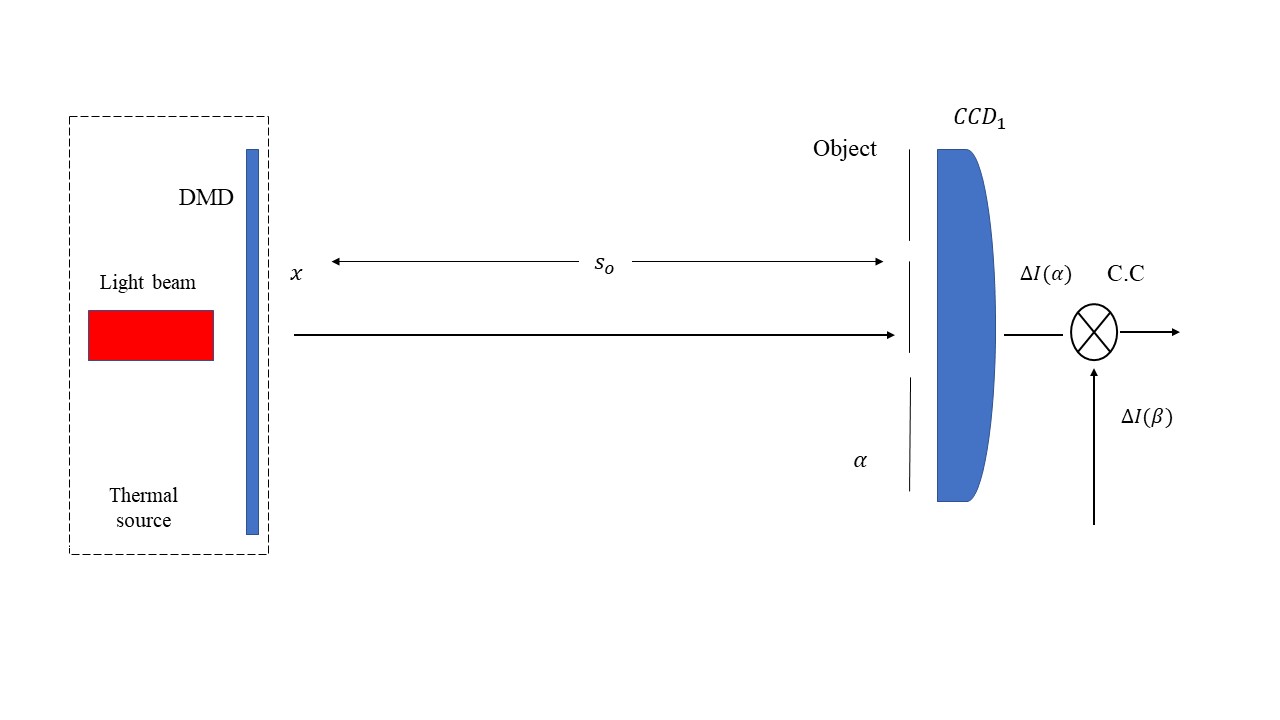}}
  \centering
  \caption{(a) Schematic diagram of the experimental setup; (b) Schematic of simplified single-arm setup, which is equivalent to (a);}\label{qvh}
  \label{fig:false-color}
\end{figure*}

The principle of the experiment is shown in Fig.~\ref{qvh}. It is similar to the traditional GI system\cite{chen2017sub}, but each beam contains a spatially resolving charge-coupled device (CCD) detector $CCD_{i}$ $(i=1,2)$ in the system. $x$, $\alpha$ and $\beta$ are the transverse coordinates on the source plane, object plane and imaging plane, respectively. The distance from the source to the object plane and image plane are $s_{o}$ and $s_{i}$, respectively.\par
In this system, we consisder that the source is monochromatic light with a wavelength $\lambda$. The light field PSF from source to object and from source to $CCD_{2}$ are writen as
\begin{equation}\label{(1)}
h_{t}(x,\alpha)=\frac{e^{-iks_{o}}}{i\lambda s_{o}}\exp(\frac{-i\pi}{\lambda s_{o}}(x-\alpha)^{2}),
\end{equation}
\begin{equation}\label{(2)}
h_{r}(x,\beta)=\frac{e^{-iks_{i}}}{i\lambda s_{i}}\exp(\frac{-i\pi}{\lambda s_{i}}(x-\beta)^{2}),
\end{equation}
where $k=\frac{2\pi}{\lambda}$. The light intensity distribution on the object and $CCD_{2}$ plane are represented as $I(\alpha)$ and $I(\beta)$, respectively. 
The transmission function of the object is represented as $t(\alpha)$.
Then, we can get that the correlation between the intensity fluctuations at two detectors is 
\begin{align}\label{(3)}
&\Delta G^{(2)}(\alpha,\beta)=\langle [I(\alpha)|t(\alpha)|^{2}-\langle I(\alpha)|t(\alpha)|^{2}\rangle][I(\beta)-\langle I(\beta)\rangle]\rangle\notag \\
&=\langle \Delta I(\alpha)|t(\alpha)|^{2}\Delta I(\beta)\rangle \notag \\
&=|\int\ G^{(1)}(x,x^{'})t(\alpha)h_{t}(x,\alpha)h^{*}_{r}(x^{'},\beta)\,dxdx^{'}|^{2}, 
\end{align}
where $\langle \dots\rangle$ represents the ensemble average. 
We consider a situation where light comes as a point-like source which is randomly and uniformly distributed on the source plane, then
\begin{equation}\label{(4)}
G^{(1)}(x,x^{'})=I_{0}\delta (x-x^{'}),
\end{equation}
where $I_{0}$ is the intensity of the source and $\delta(x)$ is the Dirac delta function. We further consider that $s_{o}=s_{i}=z$. Then, substituting Eqs.~(\ref{(1)}-\ref{(2)}) and Eq.~(\ref{(4)}) into Eq.~(\ref{(3)}), after some calculations, we have
\begin{align}\label{(5)}
\Delta G^{(2)}(\alpha,\beta)&=\langle \Delta I(\alpha)|t(\alpha)|^{2}\Delta I(\beta)\rangle \notag \\
&=I_{0}^{2}\times|t(\alpha)|^{2}\text{sinc}^{2}(\frac{2\pi R}{\lambda z} (\alpha-\beta)).
\end{align}
We integral the all $\Delta I(\alpha)|t(\alpha)|^{2}$ by $CCD_{1}$, the corresponding ghost image is given by  
\begin{equation}\label{(6)}
\Delta G^{(2)}(\beta)=\langle \Delta I(\beta)\Delta B_{t}\rangle =I_{0}^{2} \int\ |t(\alpha)|^{2}\text{sinc}^{2}(\frac{2\pi R}{\lambda z} (\alpha-\beta))\,d\alpha,
\end{equation}
where $\Delta B_{t}=\int \Delta I(\alpha)|t(\alpha)|^{2}d\alpha$, i.e., the intensity fluctuation at the bucket detector of GI system.\par
\begin{figure}[h]
  \centering
  \includegraphics[width=7 cm,height=5 cm]{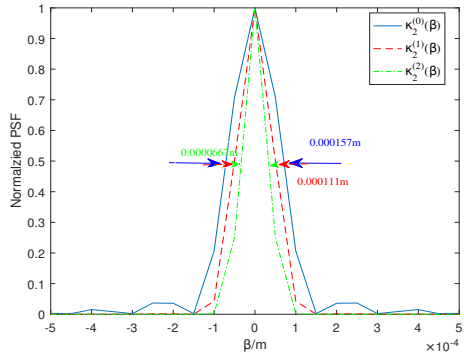}\\
  \caption{PSF of a pinhole object at $\alpha=0$ and $\lambda = 628nm, z=0.25m$ and $R=1mm$ by $\kappa^{(0)}_{2}(\beta)$ (blue solid curve), $\kappa^{(1)}_{2}(\beta)$ (ed dash cure) and $\kappa^{(2)}_{2}(\beta)$ (green dotted cure).}\label{2}
  \label{fig:false-color}
\end{figure}

If we consider that the power of the source cannot be kept stable, the $\Delta G^{(2)}(\alpha,\beta)$ in Eq.~(\ref{(5)}) and $\Delta G^{(2)}(\beta)$ in Eq.~(\ref{(6)}) should be substituted with $\Delta G^{(2)}(I_{0},\alpha,\beta)$ and $\Delta G^{(2)}(I_{0},\beta)$, respectively. Fluctuations of $I_{0}$ lead to fluctuations of $\Delta G^{(2)}(I_{0},\alpha,\beta)$ and $\Delta G^{(2)}(I_{0},\beta)$\cite{zhao2022second}. We use cumulants to describe their fluctuations information. For $\Delta G^{(2)}(I_{0},\alpha,\beta)$, the cumulant-generating function $K(s, \alpha, \beta)$ is defined as
\begin{align}\label{(7)}
K(s, \alpha, \beta)&=\ln (\langle \exp(s\Delta G^{(2)}(I_{0},\alpha,\beta))\rangle)=\sum_{n=1}^{\infty}{\kappa_{n}(\alpha, \beta)\frac{s^{n}}{n!}}\notag \\
&=\mu(\alpha,\beta)\times s+\sigma^{2}(\alpha,\beta)\times \frac{s^{2}}{2}+\dots,
\end{align}
where $\kappa_{n}(\alpha,\beta)$ is the nth-order cumulants of $\Delta G^{(2)}(I_{0},\alpha,\beta)$, $\mu(\alpha,\beta)=\langle\Delta G^{(2)}(I_{0},\alpha,\beta)\rangle$, and $\sigma^{2}(\alpha,\beta)=\langle [\Delta G^{(2)}(I_{0},\alpha,\beta)-\mu(\alpha,\beta)]^{2}\rangle$. The nth-order cumulants is given by
\begin{equation}\label{(8)}
\kappa_{n}(\alpha,\beta)=\frac{d^{(n)}K(s,\alpha, \beta)}{ds^{(n)}}|_{s=0}.
\end{equation}
From Eqs.~(\ref{(7)}-\ref{(8)}), we can get the second-order cumulant of $\Delta G^{(2)}(I_{0},\alpha,\beta)$ as
\begin{align}\label{(9)}
\kappa_{2}(\alpha,\beta)&=\langle [\Delta G^{(2)}(I_{0},\alpha,\beta)-\langle\Delta G^{(2)}(I_{0},\alpha,\beta)\rangle]^{2} \rangle\notag \\
&=\langle (I_{0}-\langle I_{0}\rangle)^{2}\rangle\times|t(\alpha)|^{4}\text{sinc}^{4}(\frac{2\pi R}{\lambda z} (\alpha-\beta)).
\end{align}
From Eq.~(\ref{(9)}), we can get that $\kappa_{2}(\alpha,\beta)$ is the fluctuation information of $\Delta G^{(2)}(I_{0},\alpha,\beta)$. Compare Eq.~(\ref{(5)}) with Eq.~(\ref{(9)}), we also can get that it has narrower PSF than that of $\Delta G^{(2)}(I_{0},\alpha,\beta)$. Then, we integral the all $\kappa_{2}(\alpha,\beta)$ by the $CCD_{1}$, and get the new imaging information as
\begin{align}\label{(10)}
&\kappa^{(1)}_{2}(\beta)=\int\kappa_{2}(\alpha,\beta)d\alpha \notag \\
&=\langle (I_{0}-\langle I_{0}\rangle)^{2}\rangle\times\int |t(\alpha)|^{4} \text{sinc}^{4}(\frac{2\pi R}{\lambda z} (\alpha-\beta))d\alpha.  
\end{align}
Compare to traditional GI in Eq.~(\ref{(6)}), Eq.~(\ref{(10)}) use $\kappa^{(1)}_{2}(\beta)$ instead of $\Delta G^{(2)}(I_{0},\beta)$ to get the information of the object, and has the narrower PSF. So, it has the better resolution than traditional GI. From Eq.~(\ref{(9)}), we also can get that the fluctuations of $I_{0}$ lead to fluctuations of $\kappa_{2}(\alpha,\beta)$. Similarly, we can further enhance the resolution of GI by the integral fluctuation information of $\kappa_{2}(\alpha,\beta)$. It is written as 
\begin{align}\label{(11)}
\kappa^{(2)}_{2}(\beta)&=\int\frac{d^{(2)}\ln (\langle \exp(s\kappa_{2}(\alpha,\beta))\rangle)}{ds^{(2)}}|_{s=0}d\alpha\notag \\
&\varpropto \int |t(\alpha)|^{8} \text{sinc}^{8}(\frac{2\pi R}{\lambda z} (\alpha-\beta))d\alpha.
\end{align}
If we use $\kappa^{(0)}_{2}(\alpha,\beta)$, $\kappa^{(0)}_{2}(\beta)$ and $\kappa^{(1)}_{2}(\alpha,\beta)$ to represent $\Delta G^{(2)}(I_{0},\alpha,\beta)$, $\Delta G^{(2)}(I_{0},\beta)$ and $\kappa_{2}(\alpha,\beta)$, respectively, according to Eq.~(\ref{(6)}) and Eqs.~(\ref{(10)}-\ref{(11)}), we can get 
\begin{align}\label{(12)}
&\kappa^{(n)}_{2}(\beta)=\int\frac{d^{(2)}\ln (\langle \exp(s\kappa^{(n-1)}_{2}(\alpha,\beta))\rangle)}{ds^{(2)}}|_{s=0}+\delta(n)[\kappa^{(0)}_{2}(\alpha,\beta)\notag \\
&-\frac{d^{(2)}\ln (\langle \exp(s\kappa^{(n-1)}_{2}(\alpha,\beta))\rangle)}{ds^{(2)}}|_{s=0}]d\alpha=\int\kappa^{(n)}_{2}(\alpha,\beta)d\alpha,
\end{align}
where $n$ is iteration times. Each iteration means that we go one step further to extract the fluctuation information. From Eq.~(\ref{(12)}), we can get that $\kappa^{(n)}_{2}(\beta)$ is the information consisting of all $\kappa^{(n)}_{2}(\alpha, \beta)$. $\kappa^{(n)}_{2}(\alpha, \beta)$ is the fluctuation information of $\kappa^{(n-1)}_{2}(\alpha, \beta)$ when $n\geqslant 1$. It is equivalent to $\Delta G^{(2)}(I_{0},\alpha,\beta)$ when $n=0$. After calculation Eq.~(\ref{(12)}), we can get that the PSF of  $\kappa^{(n)}_{2}(\beta)$ will becomes narrower as $n$ grows. In order to address a concrete example, we set $\lambda=628nm$, $z=0.25m$ and $R=1mm$. For a pinhole-like object at $\alpha=0$, the imaging results by Eq.~(\ref{(12)}) (here, we set $n$=1,2,3) are shown in Fig.~\ref{2}.
The physical significance is that when the $\kappa^{(n-1)}_{2}(\alpha,\beta)$ (here $n\geqslant 1$) has the image information about $\alpha$ points on the object plane. Its fluctuation information not only has the information about $\alpha$ points, but also has the narrower PSF than that of the original information. So, $\kappa^{(n)}_{2}(\beta)$ has the better resolution than that of $\kappa^{(n-1)}_{2}(\beta)$.\par
Due to $\kappa^{(n)}_{2}(\alpha,\beta)$ is the fluctuation information of $\kappa^{(n-1)}_{2}(\alpha,\beta)$, the signal strength of $\kappa^{(n)}_{2}(\alpha,\beta)$ is weaker than that of $\kappa^{(n-1)}_{2}(\alpha,\beta)$. So, although increasing the $n$ of the $\kappa^{(n)}_{2}(\beta)$ leads to the decrease of the FWHM of the PSF, it also results in the decrease of the signal strength.  We can set $n$ according to imaging requirements. The scheme which used $\kappa^{(n)}_{2}(\beta)$ to get the image of the object, is called iterative fluctuation ghost imaging (IFGI). It is equivalent a traditional GI when $n=0$.\par

In our previous work\cite{zhao2022second}, we have presented a scheme for GI referred to as SCGI. It is shown that $\kappa_{2}(\beta)$ is the fluctuation information of $\Delta G^{(2)}(I_{0},\beta)$, and has better resolution than that of $\Delta G^{(2)}(I_{0},\beta)$. 
In \cite{zhao2022second}, we can get that $\kappa_{2}(\beta)$ not only has $\kappa^{(1)}_{2}(\beta)$, but also has a cross-information $L(\beta)$ which reduce the resolution. This is because that in the framework of SCGI, it can get the $\Delta G^{(2)}(I_{0},\beta)$ that is the correlation between the intensity fluctuations at $\beta$ point and bucket detector, but it cannot get the $\Delta G^{(2)}(I_{0},\alpha,\beta)$ that is the correlation between the intensity fluctuations at $\alpha$ point and $\beta$ point. The fluctuation information of $\Delta G^{(2)}(I_{0},\beta)$ not only include the fluctuation information of $\Delta G^{(2)}(I_{0},\alpha,\beta)$ for different $\alpha$ points, but also include the cross-information between them, which cannot distinguish. However, we used $CCD_{1}$ instead of a bucket detector in IFGI. If the distance between the object and the $CCD_{1}$ is smaller than their longitudinal coherence length, it can get the information of the $\Delta G^{(2)}(I_{0},\alpha,\beta)$. So $L(\beta)$ is disappear in IFGI. IFGI has the better resolution limit than SCGI.\par
\begin{figure}[h]
  \centering
  \includegraphics[width=7 cm,height=5 cm]{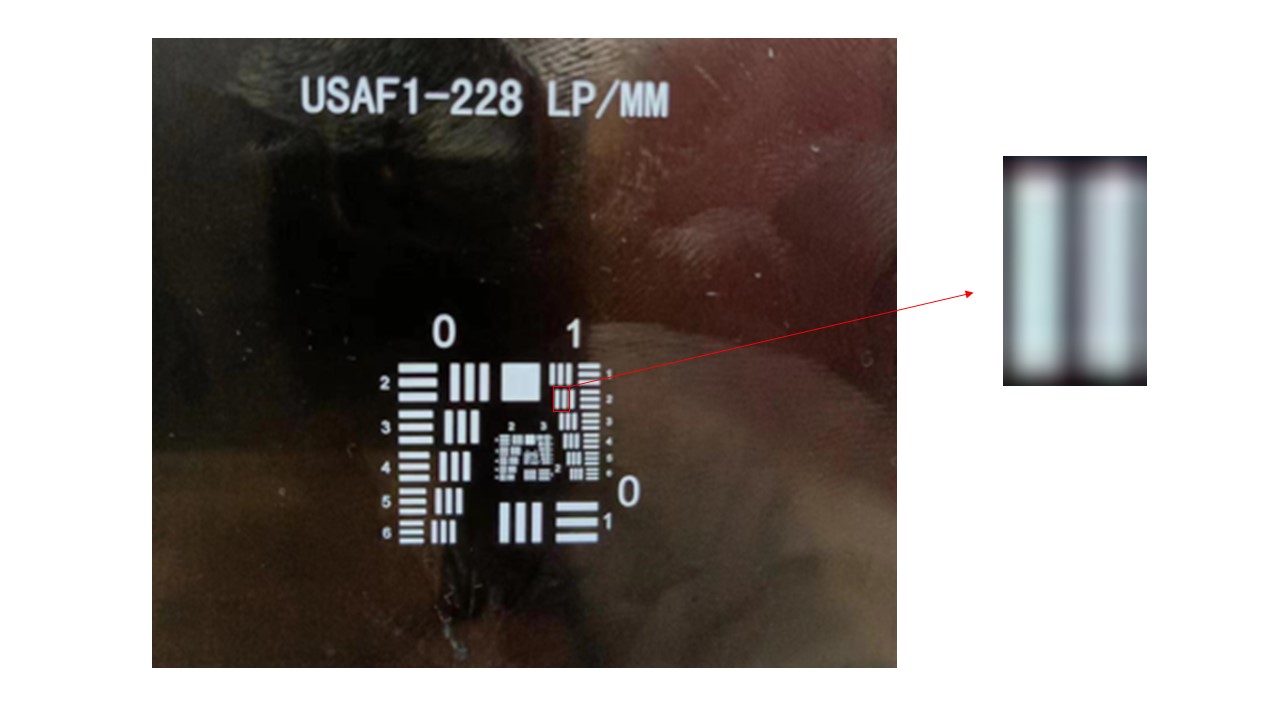}\\
  \caption{The test object for this experiment.}\label{3}
  \label{fig:false-color}
\end{figure}
\begin{figure*}[ht!]
\centering
  \subfigure[]{\includegraphics[width=4cm,height=7cm]{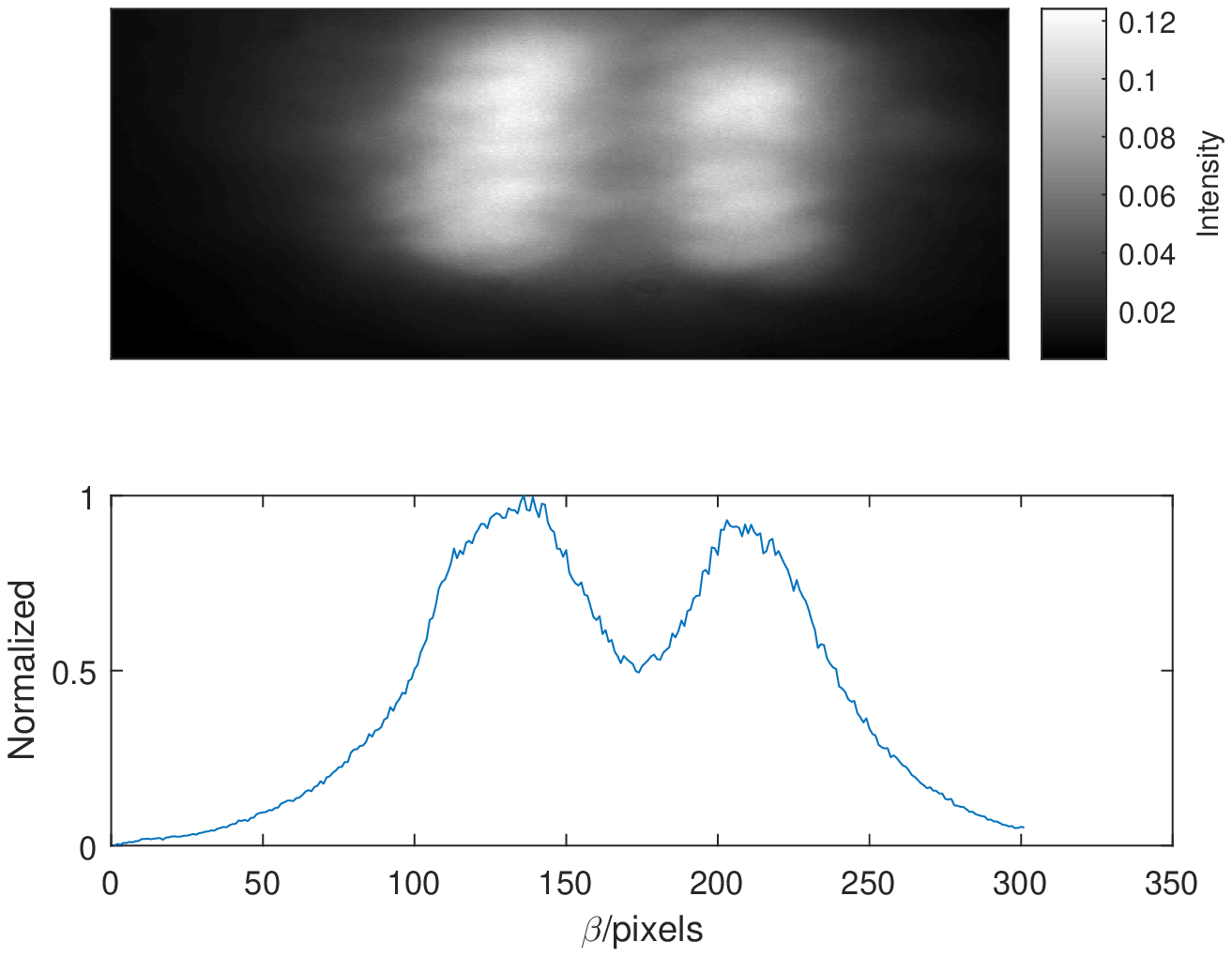}}
  \hspace{0in}
 \subfigure[]{\includegraphics[width=4cm,height=7cm]{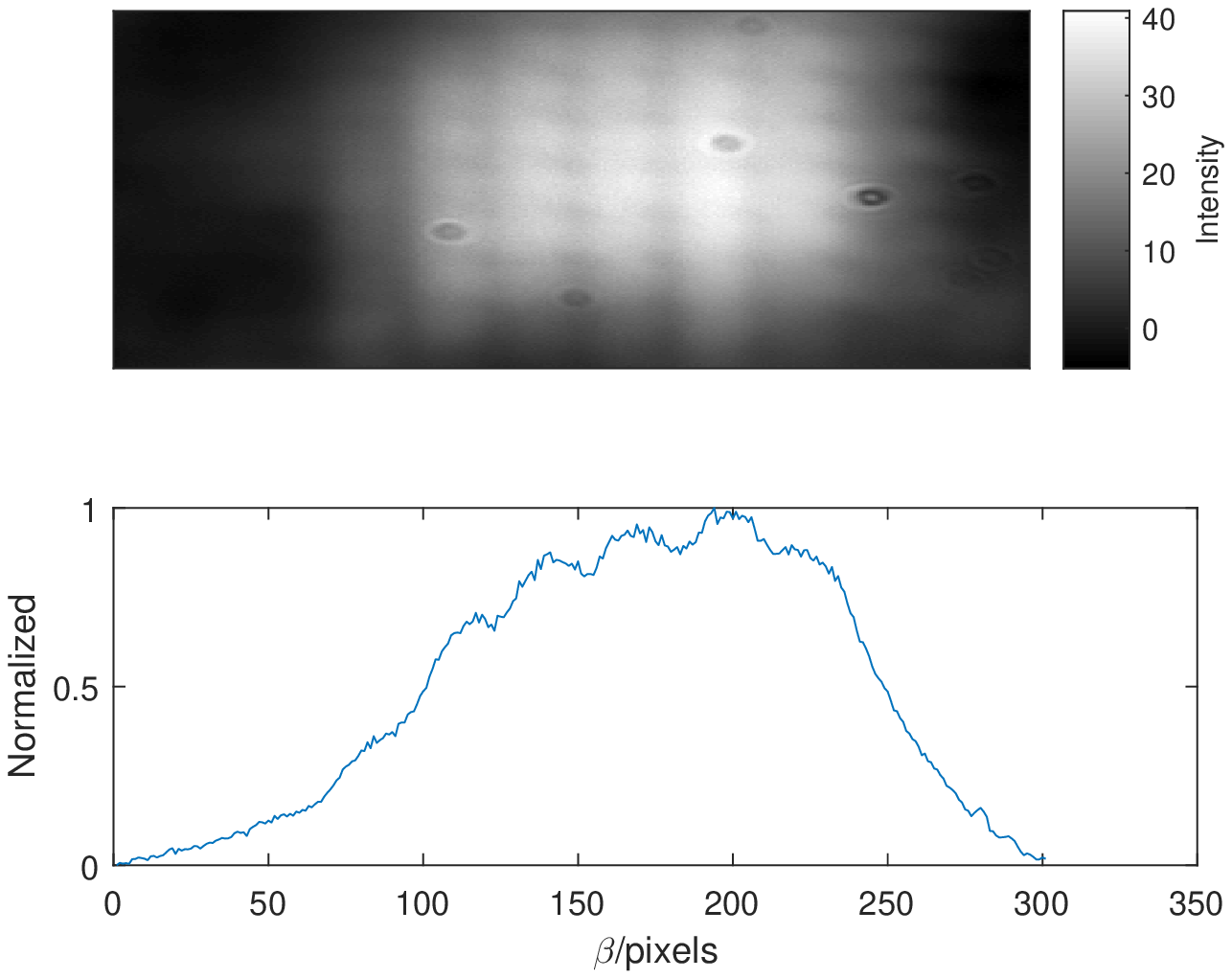}}
  \hspace{0in}
  \subfigure[]{\includegraphics[width=4cm,height=7cm]{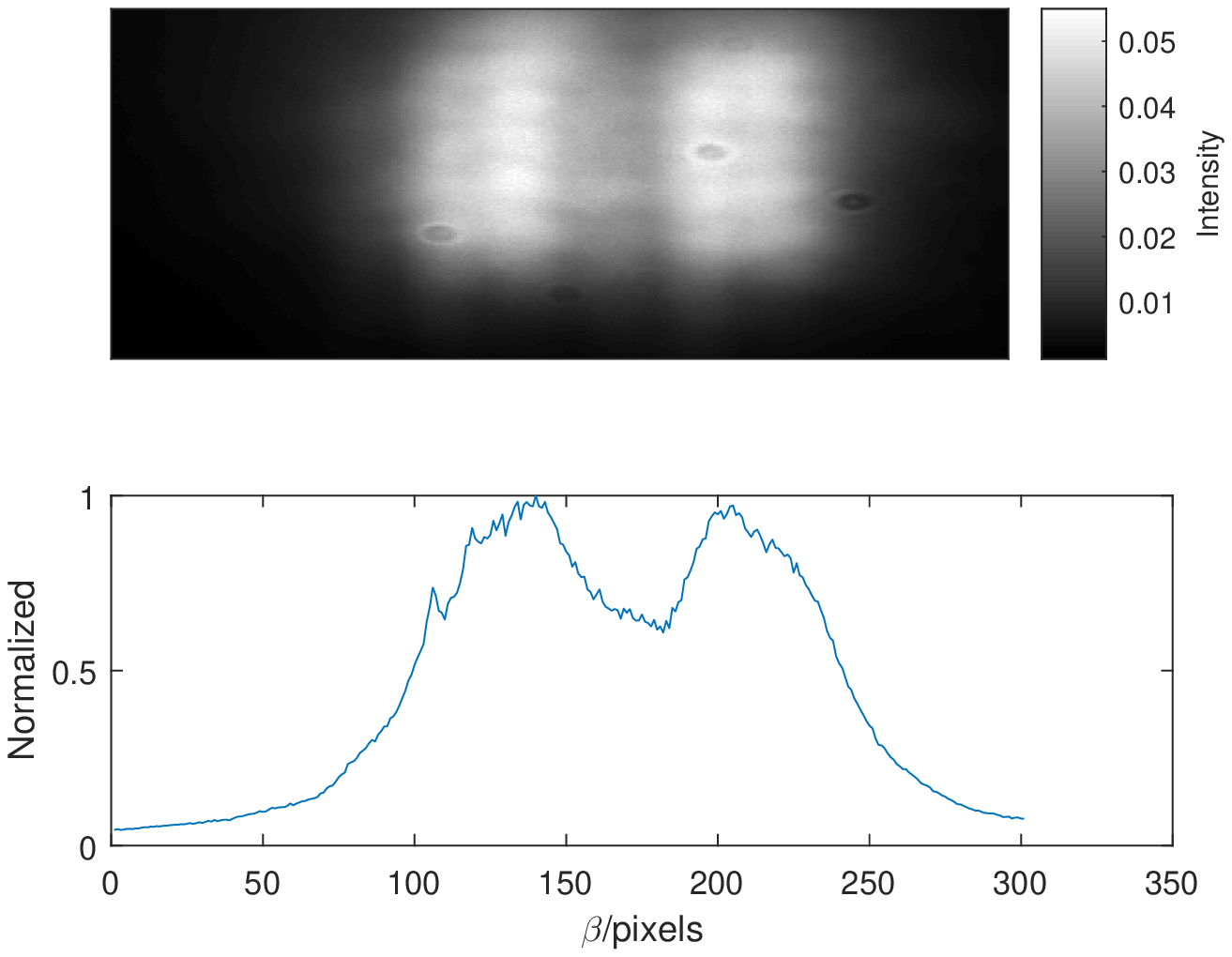}}
 \hspace{0in}
 \subfigure[]{\includegraphics[width=4cm,height=7cm]{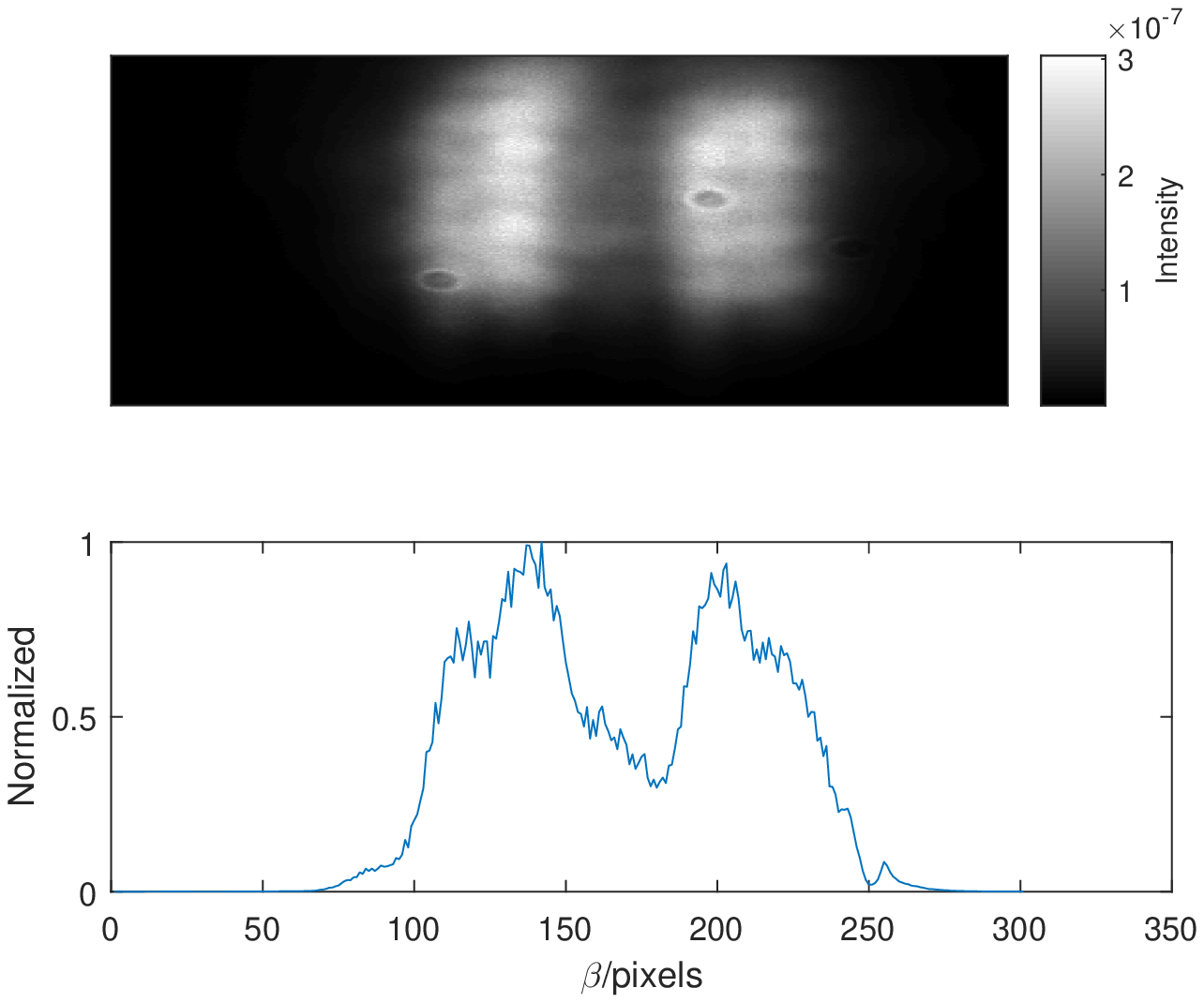}}
 \centering
  \caption{The reconstructed images of the object by $CCD_{1}$, traditional GI and IFGI; (a) $CCD_{1}$; (b)  traditional GI; (c) IFGI($\kappa^{(1)}_{2}(\beta)$); (d) IFGI ($\kappa^{(2)}_{2}(\beta)$);}\label{4}
  \label{fig:false-color}
 \end{figure*}

To verify our theoretical results, experiment results are carried out. We use the simple single-arm setup shown in Fig.~\ref{qvh}(b) (it is equivalent to Fig.~\ref{qvh}(a)\cite{chen2017sub}). The light source is a projector (XE11F), and there is a digital mirror device (DMD) in the source. An double slit with width $a=0.223mm$, slits center distance $b=0.445mm$, and slit height $g=1.114mm$ as the object. It is shown in Fig.~\ref{3}. The distance between the source plane and the object is $25cm$. A camera (MV-VEM033SM) is taken as $CCD_{1}$. The experimental process is as follows: a series of transverse patterns are generated by computer.  Each pattern is projected by a projector and illuminates the object, and the transmitted light is collected by a $CCD_{1}$. After 20000 measurements, we get 20000 group $I(\alpha)$. Then, we move the object. Above patterns are re-projected and directly received by $CCD_{1}$, respectively. We get 20000 group $I(\beta)$.
In this experiment, we use $CCD_{1}$, traditional GI and IFGI to get the image of the object, respectively. The results are shown in Fig.~\ref{4}.\par
From Figs.~\ref{4}(a-b), we can find that the resolution of the image by $CCD_{1}$ is better than that by traditional GI. This is because that the distance between the source and the object is far greater than the distance between the object and the $CCD_{1}$ in this system. From Figs.~\ref{4}(b-d), we can get that the double slit cannot be separated by traditional GI (see Fig.~\ref{4}(b)). However, it can be separated by IFGI and increasing the $n$ of $\kappa^{(n)}_{2}(\beta)$ leads to the enhancement of the resolution of IFGI 
(see Figs.~\ref{4}(c-d)). According to our previous analysis, the resolution of GI system is limited by the PSF, and  $\kappa^{(n)}_{2}(\alpha,\beta)$ has the narrower PSF than $\kappa^{(n-1)}_{2}(\alpha,\beta)$. So, $\kappa^{(n)}_{2}(\beta)$ has the better resolution than $\kappa^{(n-1)}_{2}(\beta)$. 
Compare Fig.~\ref{4}(a) with Fig.~\ref{4}(d), we can get that the resolution of IFGI is better than that by $CCD_{1}$. 
IFGI can enhance the resolution of GI via extract the fluctuation information as $n$ grows, until its resolution is close the pixel size of the detector. The experimental results agree with our theoretical analysis.\par
Compare Fig.~\ref{4}(b) with Fig.~\ref{4}(c), we can also get that the signal strength of traditional GI is better than that of IFGI. This is because that $\kappa^{(1)}_{2}(\alpha,\beta)$ is the fluctuation information of  $\Delta G^{(2)}(I_{0},\alpha, \beta)$. Compare Fig.~\ref{4}(c) with Fig.~\ref{4}(d), we also get that the signal strength of $\kappa^{(2)}_{2}(\beta)$ is weaker than $\kappa^{(1)}_{2}(\beta)$. This because that $\kappa^{(n)}_{2}(\alpha,\beta)$ is the fluctuation information of $\kappa^{(n-1)}_{2}(\alpha,\beta)$ when $n\geqslant 1$. Above experimental results proved that the resolution of IFGI become better and the signal strength become weaker as $n$ grows. In any case, we can have a tradeoff according to the image requirements.\par
In conclusion, we have presented a technique called IFGI that enhances dramatically the resolution of the GI via the fluctuation characteristics of the second-order correlation function. It is found that the resolution is strongly dependent on the iteration times $n$ when the pixel size of the detector is much smaller than the average size of the speckles. Although a trade-off between resolution and signal strength has to be considered, we select $n$ according to the image requirements. It offers a general and alternative approach applicable to all fields of imaging where higher resolution is needed.\par

 \section*{Acknowledgement}
This work is supported by the Science \& Technology Development Project of Jilin Province (No.YDZJ202101ZYTS030).\par
\section*{Disclosures}
\noindent\textbf{Disclosures.} The authors declare no conflicts of interest.
\bibliography{article1}
\bibliographyfullrefs{article1}

\end{document}